\documentclass[reprint,aps,amsmath,amssymb,nofootinbib,superscriptaddress]{revtex4-2}

\usepackage{mathtools}
\usepackage{mathrsfs} 
\usepackage{braket} 
\usepackage[dvipsnames]{xcolor}  
\usepackage{float}
\usepackage[T1]{fontenc}
\usepackage[utf8]{inputenc}
\usepackage{graphicx,color,rotating,pifont}
\usepackage{amsmath,amssymb}
\usepackage{mathtools}
\usepackage{subfigure}
\usepackage{siunitx}
\usepackage{bm}
\usepackage{ae}
\usepackage{dcolumn}
\usepackage{txfonts}
\usepackage{tensor}
\usepackage{braket}
\usepackage{booktabs}
\usepackage{xspace} 
\usepackage{soul}
\usepackage{multirow}
\setstcolor{red}
\setul{}{.2ex} 

\usepackage[normalem]{ulem}

\usepackage{tikz}
\usetikzlibrary{positioning,calc,shapes,decorations.markings,decorations.pathmorphing}
\tikzset{asg/.cd,
  omega-vertex/.style={circle,solid,draw=black,fill=white,minimum size=5pt, inner sep=0pt},
  dbd-vertex/.style={coordinate},
  pline/.style={thick, postaction={decorate}, decoration={markings, mark=at position .5 with {\arrow[xshift=2pt]{stealth}}}},
  hline/.style={thick, postaction={decorate}, decoration={markings, mark=at position .5 with {\arrowreversed[xshift=-2pt]{stealth}}}},
  shift arrow/.style={/pgf/decoration/transform={xshift=#1}},
  shift arrow/.default=-2pt,
  dbd-2b/.style={decorate, decoration=snake},
  omega-2b/.style={densely dashed},
  neutron/.style={draw=blue},
  proton/.style={draw=red},
}

\usepackage[colorlinks,linkcolor=blue,anchorcolor=blue,citecolor=blue]{hyperref} 
\hypersetup{
   colorlinks=true, 
   linkcolor=blue,  
   citecolor=blue,  
   filecolor=blue,  
   urlcolor=blue    
} 

\DeclareMathSymbol{\NS}{\mathord}{AMSb}{"4E}

\DeclareSIUnit{\fm}{\femto\meter}

\newcommand{\abs}[1]{\lvert {#1} \rvert}

\newcommand{\beq}{\begin{equation}}
\newcommand{\eeq}{\end{equation}}
\newcommand{\beqn}{\begin{eqnarray}}
\newcommand{\eeqn}{\end{eqnarray}}
\newcommand{\bsub}{\begin{subequations}}
\newcommand{\esub}{\end{subequations}}
\newcommand{\bpm}{\begin{pmatrix}}
\newcommand{\epm}{\end{pmatrix}}

\newcommand\identity{1\kern-0.25em\text{l}}

\begin{document}

\title{Benchmarking rotational correction energies in odd-mass nuclei}

  \author{Y. Li }    
  \affiliation{School of Physics and Astronomy, Sun Yat-sen University, Zhuhai 519082, P.R. China}  

    \author{ E. F. Zhou}    
  \affiliation{School of Physics and Astronomy, Sun Yat-sen University, Zhuhai 519082, P.R. China}

    \author{J. M. Yao }    
  \email{Corresponding author: yaojm8@sysu.edu.cn}
  \affiliation{School of Physics and Astronomy, Sun Yat-sen University, Zhuhai 519082, P.R. China}  
  
\date{\today}

\begin{abstract} 
Nuclear energy density functional theory (DFT) provides a microscopic approach to describing nuclear masses. By incorporating pairing correlations and deformation within DFT, nuclear masses can be predicted with sub-MeV accuracy. A crucial factor in achieving this precision is the dynamical correlation energy associated with restoring rotational symmetry, known as rotational correction energy (RCE). This correction is typically estimated using the cranking approximation in perturbation theory. In this work, we benchmark the results of these calculations against exact angular-momentum projection (AMP) for both even-even and odd-mass nuclei from light to heavy mass region, utilizing a covariant DFT framework. We find that the RCE, computed using the full moment of inertia (MoI) in the cranking approximation, closely matches the results from exact AMP for both even-even and odd-mass nuclei, with the exception of near-spherical even-even nuclei. For certain configurations in odd-mass nuclei, however, the RCEs can become abnormally small, a phenomenon linked to the divergence of the MoI.   To address this,  a regulator is introduced for the MoI of an odd-mass nucleus, the validity of which is examined through exact AMP calculations. 
\end{abstract}
  
\maketitle

\section{Introduction} 
 
The nuclear mass is a fundamental property crucial for understanding the stability of nuclear many-body systems and plays a key role in uncovering the origins of elements in the Universe~\cite{Burbidge:1957RMP}. Accurate knowledge of nuclear masses is critical for describing the abundance of heavy elements produced via the rapid neutron capture process~\cite{Sun:2008PRC,Mumpower:2016,Jiang:2021}. Over the past decades, significant progress has been made in both nuclear mass measurements~\cite{Wang:2021} and the precision of nuclear mass predictions~\cite{Goriely:2009_Skyrme,Wang:2014,Niu:2018}. Among all mass models, the self-consistent mean-field implementation of density functional theory (DFT) stands out as a powerful microscopic approach capable of describing nuclei across the entire mass range, from light to heavy. This has been achieved using both non-relativistic energy density functionals (EDFs), such as Skyrme~\cite{Samyn:2003swt,Goriely:2009_Skyrme} and Gogny~\cite{Goriely:2009_Gogny} forces, as well as relativistic covariant density functional theory (CDFT)~\cite{Geng:2005,Afanasjev:2016,Pena-Arteaga:2016clz,DRHBcMassTable:2022uhi,DRHBcMassTable:2024}. In all these models, dynamical correlation energies (DCEs), including rotational and vibrational corrections, have proven essential for improving the accuracy of nuclear mass descriptions~\cite{Goriely:2010bm,Chamel:2008aa,Samyn:2004bm,Zhao:2012gn,Meng:2013pa}.

The rotational correction energy (RCE) arises when the DFT yields a deformed nuclear solution, breaking rotational symmetry~\cite{Ring:1980}. Restoring this symmetry results in an energy gain of several MeV~\cite{Yao:2022HBNP}. Similarly, the vibrational correction energy emerges from quantum fluctuations around the equilibrium shape. Both effects can be accurately accounted for within the quantum-number projected generator coordinate method (PGCM)\cite{Sheikh:2021qv}. However, the computational complexity of PGCM makes its application across the entire nuclear chart challenging, though it has been achieved for all even-even nuclei with known binding energies using Skyrme\cite{Bender:2004Global} and Gogny~\cite{Rodriguez:2014Global} forces with certain approximations. Alternatively, these DCEs are typically calculated at the mean-field level using a parameterized formula~\cite{Chamel:2008aa,Goriely:2010bm}
\begin{equation}
\label{eq:pheno_formulas_cranking}
  E_{\rm DCE} =\frac{\braket{\hat J^2}}{2\mathcal{I}_{\rm IB}}\left\{ b \tanh(c\abs{\beta_2}) + d\abs{\beta_2}
  \exp\left[-l(\abs{\beta_2}-\beta_2^0)^2 \right] \right\}, 
\end{equation} 
where the first term in the brace bracket represents the RCE, and the second accounts for the deformation dependence of the vibrational correction energy. The parameter $\beta_2$ represents the quadrupole deformation of the mean-field solution, while $b$, $c$, $d$, $l$, and $\beta_2^0$ are free parameters fitted to nuclear mass data. The moment of inertia (MoI), $\mathcal{I}_{\rm IB}$, is calculated using the Inglis-Belyaev formula~\cite{belyaev1959}, with $\hat{J}^2$ as the squared angular momentum operator. In the CDFT description of even-even nuclei from $Z=8$ to $Z=108$ based on the PC-PK1 EDF\cite{Zhao:2010PRC}, incorporating the DCEs of (\ref{eq:pheno_formulas_cranking}) reduces the root-mean-square (rms) error in nuclear mass predictions from 2.58 MeV to 1.24 MeV\cite{Zhang:2013FP}. Similar improvements were observed when DCEs were considered using a five-dimensional collective Hamiltonian~\cite{Lu:2015PRC,Sun:2022CPC}.

To date, the precision of nuclear mass predictions in state-of-the-art EDF studies has reached the sub-MeV level~\cite{Goriely:2009_Skyrme,Niu:2018,Scamps:2020fyu}, which is comparable to the uncertainties introduced by different methods for  DCEs. Therefore, it is both timely and crucial to validate the formula (\ref{eq:pheno_formulas_cranking}) against PGCM calculations, particularly for odd-mass nuclei, where dynamical correlations have been less thoroughly explored. Accurate calculations of RCEs for both even-even and odd-mass nuclei can significantly impact predictions of odd-even mass differences and nucleon separation energies~\cite{Rodriguez-Guzman:2014,Wu:2019PRC}. Furthermore, RCEs could potentially modify the energy ordering of states with different shapes in nuclei exhibiting shape coexistence~\cite{Fu2013:PRC}, thereby influencing the understanding of low-energy nuclear structure. Recently, we extended the PGCM framework based on CDFT—referred to as multi-reference (MR)-CDFT~\cite{Yao:2009PRC,Yao:2014}—to odd-mass nuclei~\cite{Zhou:2023odj}. This advancement enables a benchmark study of RCEs in both even-even and odd-mass systems, providing valuable insights into the accuracy of these correction energies.

The paper is organized as follows. In Sec.~\ref{sec:formalism}, we present the formulas for the RCEs from the exact angular-momentum projection (AMP) calculation as well as those derived using the cranking approximation in perturbation theory. We also review the formulas for the MoIs of both even and odd-mass nuclei. In Sec.~\ref{sec:results}, we benchmark the RCE results obtained from the cranking approximation against the exact AMP calculation and examine the impact of RCEs on neutron separation energies and nuclear masses. Finally, we summarize the main findings of this work in Sec.~\ref{sec:summary}.

\section{Theoretical framework} 
\label{sec:formalism}

\subsection{Rotational correction energy in the AMP}

Since the RCE, gained from the restoration of rotational symmetry, enhances nuclear binding energy, we define the RCE from the exact AMP calculation for the lowest-energy mean-field state $\ket{\Phi_0}$, projected onto neutron and proton numbers $(N, Z)$, as follows:
\begin{equation} 
\label{eq:PNAMP_rot} 
E^{\rm PNAMP}_{\rm rot} =\frac{\bra{\Phi_0}\hat{H}_0\hat{P}^{N}\hat{P}^{Z} \ket{\Phi_0}}{\bra{\Phi_0} \hat{P}^{N}\hat{P}^{Z}\ket{\Phi_0}} -\frac{\bra{\Phi_0}\hat{H}_0\hat{P}^{J}_{KK}\hat{P}^{N}\hat{P}^{Z}\ket{\Phi_0}}{\bra{\Phi_0} \hat{P}^{J}_{KK}\hat{P}^{N}\hat{P}^{Z}\ket{\Phi_0}}, \end{equation}
where $\hat{P}^{J}_{KK}$ and $\hat{P}^{N, Z}$ are projection operators that select components with angular momentum $J$ and its $z$-component $K$, neutron number $N$, and proton number $Z$~\cite{Ring:1980}. These operators are defined as follows:

\begin{subequations} 
\begin{align} 
\hat P^{J}_{KK} &= \frac{2J+1}{8\pi^2}\int d\Omega D^{J\ast}_{KK}(\Omega) \hat{R}(\Omega), \\  
\hat{P}^{N_\tau} &= \frac{1}{2\pi}\int^{2\pi}_0 d\varphi_{\tau} e^{i\varphi_{\tau}(\hat{N}_\tau-N_\tau)}. 
\end{align} 
\end{subequations}
The Wigner $D$-function, $D^{J}_{KK}(\Omega)=\bra{JK}\hat{R}(\Omega)\ket{JK}$, depends on the Euler angles $\Omega=(\phi, \theta, \psi)$. The mean-field wave function $\ket{\Phi_0}$ is obtained using CDFT  with axial symmetry employing the PC-F1 parameterization~\cite{Burvenich:2002PRC}, combined with Bardeen-Cooper-Schrieffer (BCS) theory. The use of other parameterization, such as PC-PK1~\cite{Zhao:2010PRC} does not change the conclusion of this work. The auxiliary single-particle wave functions for neutrons and protons are described by Dirac spinors, which are solved self-consistently by expanding their large and small components in a spherical harmonic oscillator (HO) basis. The oscillator frequency is set to $\hbar\omega_0 = 41A^{-1/3}$ MeV, where $A$ is the mass number.

Some additional points should be noted regarding the mean-field wave function.
\begin{itemize}
    \item For an even-even nucleus, the spin-parity of the ground state is always $J^\pi=0^+$. The mean-field state $\ket{\Phi_0}$ is approximated as a quasiparticle vacuum and it is constructed as~\cite{Ring:1980} 
\begin{equation}
    \ket{\Phi_0} =  \prod_{k>0}(u_k + \varv_k a_{k}^{\dagger}a_{\bar{k}}^{\dagger})\ket{0}
\end{equation}
where $u^2_k=1-\varv^2_k$ with $\varv^2_k$ being the occupation probability of the $k$-th single-particle state, determined by the BCS theory.

\item  For an odd-mass nucleus,  the mean-field wave function $\ket{\Phi_0}$ is approximated by the lowest-energy one-quasiparticle state with the quantum numbers $K^\pi$ determined by the blocked orbital,
\begin{equation}
\label{eq:odd-mass-wf}
     \ket{\Phi_0^{(\rm OA)}(k_b) }  =  \alpha_{k_b}^{\dagger} \ket{\Phi_0 (k_b)},
\end{equation}
 where $\ket{\Phi_0{(k_b)}}$ is a quasiparticle vacuum constructed as follows~\cite{Zhou:2023odj}
\begin{equation}
\label{eq:odd-mass-qpv}
    \ket{\Phi_0{(k_b)}} = (1+a_{k_b}^{\dagger}a_{\bar{k_b}}^{\dagger}) \prod_{k\neq k_b}(u_k + \varv_k a_{k}^{\dagger}a_{\bar{k}}^{\dagger})\ket{0}.
\end{equation}
In this approximation, the time-reversal partner of the blocked $k_b$-th single-particle state is half-occupied.
In this work, we assume that the spin-parity $J^\pi$ of the ground state is equal to the quantum numbers  $K^\pi$ of the blocked single-particle state which is a reasonable assumption for deformed states.

\end{itemize}

Alternatively, the RCE can also be defined for the mean-field state $\ket{\Phi_0}$ without the particle-number projection (PNP),
\beq
\label{eq:AMP_rot}
E^{\rm AMP}_{\rm rot} 
=\bra{\Phi_0}\hat{H_0} \ket{\Phi_0}-\frac{\bra{\Phi_0}\hat{H_0}\hat{P}^{J}_{KK} \ket{\Phi_0}}{\bra{\Phi_0} \hat{P}^{J}_{KK} \ket{\Phi_0}}.
\eeq  
Obsviously, the RCEs defined by both $E^{\rm PNAMP}_{\rm rot}$ and $E^{\rm AMP}_{\rm rot}$ are positve. We will compare the values of the above two definitions for a set of odd-mass nuclei.
More details about the projection calculation for even-even and odd-mass nuclei based on the CDFT can be found, for instance, in Refs.~\cite{Yao:2009PRC,Zhou:2001su}.

\subsection{Rotational correction energy in the cranking approximation }
The energy of angular-momentum projected state can be approximately determined using  the Kamlah expansion~\cite{Kamlah:1968} up to second order~\cite{Ring:1980}. In the cranking approximation with the constraint $\braket{\hat{J}_x}=\sqrt{J(J+1)}$, the energy of the projected state is simplified as~\cite{Beck1970,Hara:1982NPA}
\beq
 E^J_{\rm AMP}=\frac{
\bra{\Phi_0}\hat{H_0}\hat{P}^{J}_{KK}\ket{\Phi_0}}{\bra{\Phi_0} \hat{P}^{J}_{KK}\ket{\Phi_0}} 
\simeq  \bra{\Phi_0}\hat{H_0} \ket{\Phi_0}
-E_{\rm rot},
\eeq
where the RCE is approximated as
\beq
\label{eq:RCE_cranking}
E_{\rm rot}
=\frac{\bra{\Phi_0}\Delta\hat{J}^2\ket{\Phi_0}}{2\mathcal{I}_Y}, 
\eeq
with $\Delta \hat{\mathrm{J}}= \hat{\mathrm{J}}  - \braket{\hat{ \mathrm{J}}}$ with $\braket{\cdots}$ representing the expectation value with respect to the mean-field state $\ket{\Phi_0}$, and $\braket{\Delta\hat{J}^2}=\braket{\hat{J}^2} -\braket{\hat{J}}^2$. For the ground states of even-even nuclei, $\braket{\hat{J}}=0$, while for odd-mass nuclei, $\braket{\hat{J}}=K$.  The expectation value of $\hat J^2$ is determined by
\beqn 
\label{eq:expect_J2}
\braket{\hat{J}^2} 
&=& \sum_{i=x,y,z}\sum_{m_1m_2m_3m_4} J^{(i)}_{m_1m_2}J^{(i)}_{m_3m_4}  \nonumber\\
&\times& 
\Bigg( 
\rho_{m_2m_1}\rho_{m_4m_3} - \rho_{m_4m_1}\rho_{m_2m_3}
+\rho_{m_4m_1}\delta_{m_3m_2} - \kappa^{*}_{m_1m_3}\kappa_{m_4m_2}
\Bigg), \nonumber\\
\eeqn 
where $J^{(i)}_{m_1m_2}$ is the matrix element of the $i$-th component of angular momentum $\hat{\mathbf{J}}$ in the HO basis, and the density matrices in this basis are defined as 
\bsub\beqn 
\rho_{mn}  &=& \braket{c^\dagger_nc_m},\\
\kappa_{mn} &=& \braket{c_nc_m},\\
\kappa_{mn}^{*} &=& \braket{c_m^\dagger c_n^\dagger},
\eeqn 
\esub
with $c^\dagger, c$ being the single-particle creation and annihilation operators, respectively.

In practical calculations, the Yoccoz MoI $\mathcal{I}_Y$ is replaced by the MoI determined in the cranking approximation~\cite{Ring:1980},
\begin{equation}
  \label{eq:MOI}
  \mathcal{I}_{\rm Cr} = \frac{\bra{\Phi_\omega}\hat{J}_x \ket{\Phi_\omega}}{\omega}
\end{equation}
where $\ket{\Phi_{\omega}}$ is the cranked mean-field  wave function of the Routhian operator $\hat{H}_\omega= \hat{H}_0 - \omega \hat{J}_x$ and it can be obtained by the perturbation theory.  For even-even nuclei,  by truncating the wave functions of the excited states up to  two-quasiparticle configurations, the cranked wave function is given by  
\begin{equation}
   \label{eq:Phi_omega}
    \ket{\Phi_\omega } \simeq \ket{\Phi_0} + \omega\sum_{k<k^\prime}\frac{\bra{\Phi_0}\alpha_{k^\prime}\alpha_{k}\hat{J}_x\ket{\Phi_0}}{E_{k} + E_{k^\prime}}\alpha_{k}^{\dagger}\alpha_{k^\prime}^{\dagger}\ket{\Phi_0},
\end{equation}
 where $E_{k}$ and $E_{k^\prime}$ are quasiparticle energies given by 
\begin{equation}
\label{eq:quasi-particle-energies}
  E_{k} = \sqrt{(\epsilon_k - \lambda_\tau)^2 + f_k^2 \Delta_k^2}.
\end{equation}
Here $\epsilon_k$ is the energy of the $k$-th single-particle state, and $\lambda_\tau$ is the Fermi energy ($\tau = n$ for neutron and $\tau=p$ for proton). The smooth energy-dependent cutoff weights $f_k$ is introduced  to simulate the effect of finite-range pairing force and $\Delta_k$ is the single-particle pairing gap \cite{Bender_2000,Yao:2009PRC}.

Substituting (\ref{eq:Phi_omega}) into (\ref{eq:MOI}), one obtains the MoI in the cranking approximation, as originally derived by Nilsson and Prior~\cite{nilsson1961}
\beqn
  \label{eq:NP}
    \mathcal{I}_{\rm NP} 
    &=& \mathcal{I}_1 + \mathcal{I}_2 \nonumber\\
    &=& 2\sum_{k,k^{\prime}>0} \frac{| \bra{k}\hat{J}_x\ket{{k}^{\prime}}|^2 }{E_{k}+E_{k^{\prime}}}\eta^2_{kk'} 
    + 2\sum_{k,k^{\prime}>0} \frac{ |\bra{\bar{k}}\hat{J}_x\ket{{k}^{\prime}} |^2}{E_{k}+E_{k^{\prime}}}\eta^2_{kk'},
 \eeqn
where $\eta_{kk'}=u_{k}\varv_{k^{\prime}} -\varv_{k}u_{k^{\prime}} $. The first term $\mathcal{I}_1$ is just the Inglis-Belyaev MoI $\mathcal{I}_{\rm IB}$ ~\cite{belyaev1959} that has been frequently employed in the calculation of RCE~\cite{Chamel:2008aa,Goriely:2009_Skyrme}. The second term $\mathcal{I}_2$  is nonzero only in the case of $K_{k'}=K_{k} =1/2$.

Similar to (\ref{eq:Phi_omega}), the wave function $\ket{\Phi_{\omega}^{(\rm OA)}}$ of the cranked mean-field state for an odd-mass nucleus is obtained by the perturbation theory, 
\beqn
  \label{eq:PhiomegaOA} 
&&\ket{\Phi_{\omega}^{(\rm OA)}} \nonumber\\ 
&=& \ket{\Phi_0^{(\rm OA)}(k_b)} + \omega \sum_{k\neq \pm k_b} \frac{\bra{\Phi_0 {(k_b)}}\alpha_{k}\hat{J}_x\ket{\Phi_0^{\rm (OA)}(k_b)}}{E_k - E_{k_b}}\alpha_{k}^\dagger\ket{\Phi_0{(k_b)}} \nonumber \\
&& + \omega \sum_{k<k^\prime; k,k^\prime\ne k_b }\frac{\bra{\Phi_0(k_b)}\alpha_{k_b} \alpha_{k^\prime}\alpha_k\hat{J}_x\ket{\Phi_0^{\rm (OA)}(k_b)}}{E_k + E_{k^\prime}}\alpha_k^
    \dagger \alpha_{k^\prime}^\dagger\alpha_{k_b}^\dagger\ket{\Phi_0(k_b)} \nonumber \\
  \eeqn
where  the wave functions of excited states  are approximated by one- and three-quasiparticle states.

Substituting (\ref{eq:PhiomegaOA}) into (\ref{eq:MOI}), one obtains a formula for the full MoI $\mathcal{I}_{F}$ of an odd-mass nucleus~\cite{prior1968,Zhou:2001su}, which is composed of four terms,
\beqn 
  \label{eq:IF} 
        \mathcal{I}_{\rm F}
        &=&\sum_{\alpha=1}^4\mathcal{I}_{\alpha}\nonumber\\
        &=&  2\sum_{k,k^{\prime}>0;k\ne k_b} \frac{| \bra{k}\hat{J}_x\ket{{k^{\prime}}}|^2  }{E_{k}+E_{k^{\prime}}} \eta^2_{kk'}
        +2\sum_{k,k^{\prime}>0;k\ne k_b} \frac{| \bra{\bar{k}}\hat{J}_x\ket{{k^{\prime}}}|^2  }{E_{k}+E_{k^{\prime}}} \eta^2_{kk'}\nonumber\\
        &&+ 2\sum_{k>0; k \ne k_b}  \frac{  |\bra{k}\hat{J}_x\ket{{k_b}}|^2  }{E_{k}-E_{k_b}}  \xi^2_{kk_b}
        + 2\sum_{k>0; k \ne k_b}  \frac{  |\bra{\bar{k}}\hat{J}_x\ket{{k_b}}|^2  }{E_{k}-E_{k_b}}  \xi^2_{kk_b},\nonumber\\
   \eeqn
where $\xi_{kk'}=u_{k} u_{k_b} + \varv_{k}\varv_{k_b}$. The first and third terms  in the last equivalence correspond to the $\mathscr{I}_\alpha$ term (4) in Ref.~\cite{prior1968}. For convenience, we introduce the following notations,
\bsub\beqn
  \label{eq:IPBN} 
    \mathcal{I}_{\rm PBN}
    &\equiv&  \mathcal{I}_{1} + \mathcal{I}_{3}=\mathcal{I}_{1+3}, \\
    \mathcal{I}_{2+4} &\equiv&  \mathcal{I}_{2}+\mathcal{I}_4
  \eeqn
 \esub
 The  $\mathcal{I}_{2+4}$ term gives the $W_\alpha$ term (5) in Ref.~\cite{prior1968}, where the $\mathcal{I}_2$ is nonzero in the case of $K_{k'}=K_{k} =1/2$, while  the $\mathcal{I}_4$ is nonzero if $K_k=K_{k_b}=1/2$.   The  $\mathcal{I}_{3+4}$  terms could be divergent if the denominator is approaching zero. To address this issue, we regularize this term 
 \beqn
  \label{eq:regulated_MoI} 
 \mathcal{I}_{3+4} \to \mathcal{I}^R_{3+4} 
 &=& 2\sum_{k>0; k \ne k_b}  \frac{ {f(E_{k},E_{k_b})}}{E_{k}-E_{k_b}}  |\bra{k}\hat{J}_x\ket{{k_b}}|^2  \xi^2_{kk_b} \nonumber\\
 &&+ 2\sum_{k>0; k \ne k_b}  \frac{{f(E_{k},E_{k_b})}}{E_{k}-E_{k_b}} |\bra{\bar{k}}\hat{J}_x\ket{{k_b}}|^2 \xi^2_{kk_b}
 \eeqn 
by introducing the following regulator,
 \beq
  \label{eq:regulator} 
 f(E_{k}, E_{k_b})
 =1 - \frac{\exp{\Bigg[-k_0(E_{k}-E_{k_b})^4)\Bigg]}}{1+\exp{\Bigg[-k_1(\abs{\beta_2}-\beta_0 )\Bigg] }}.
 \eeq 
We will show numerically that  the regulator removes the abnormal behavior of the MoI around the pole $E_k=E_{k_b}$, keeping other part almost unchanged if the parameter $k_0$ is chosen to be 0.05 MeV$^{-4}$, and $k_1$ and $\beta_0$ are  chosen as 100 and 0.2, respectively.

\section{Results and discussions}
\label{sec:results}

\begin{figure}[tb]
	\centering
	\includegraphics[width=0.9\columnwidth]{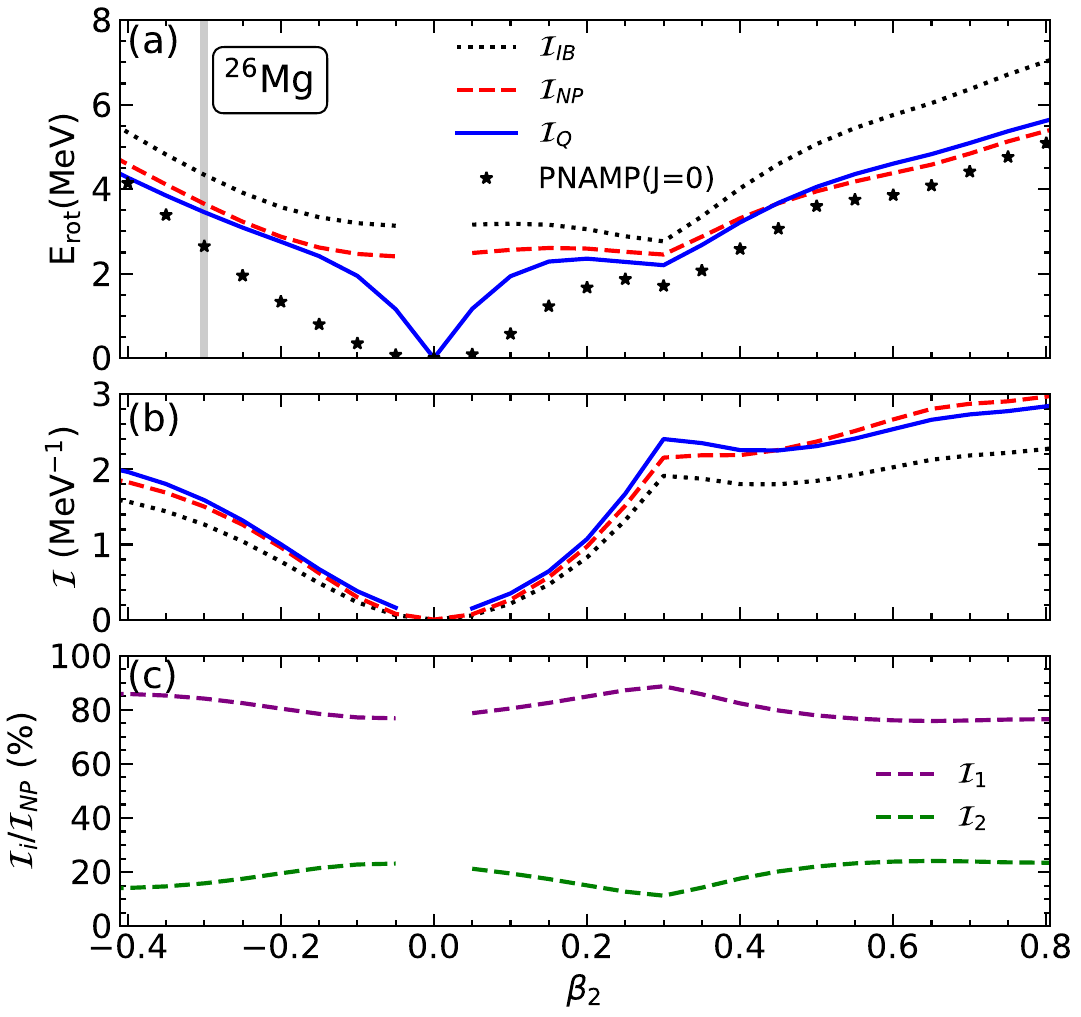}  
	\includegraphics[width=0.9\columnwidth]{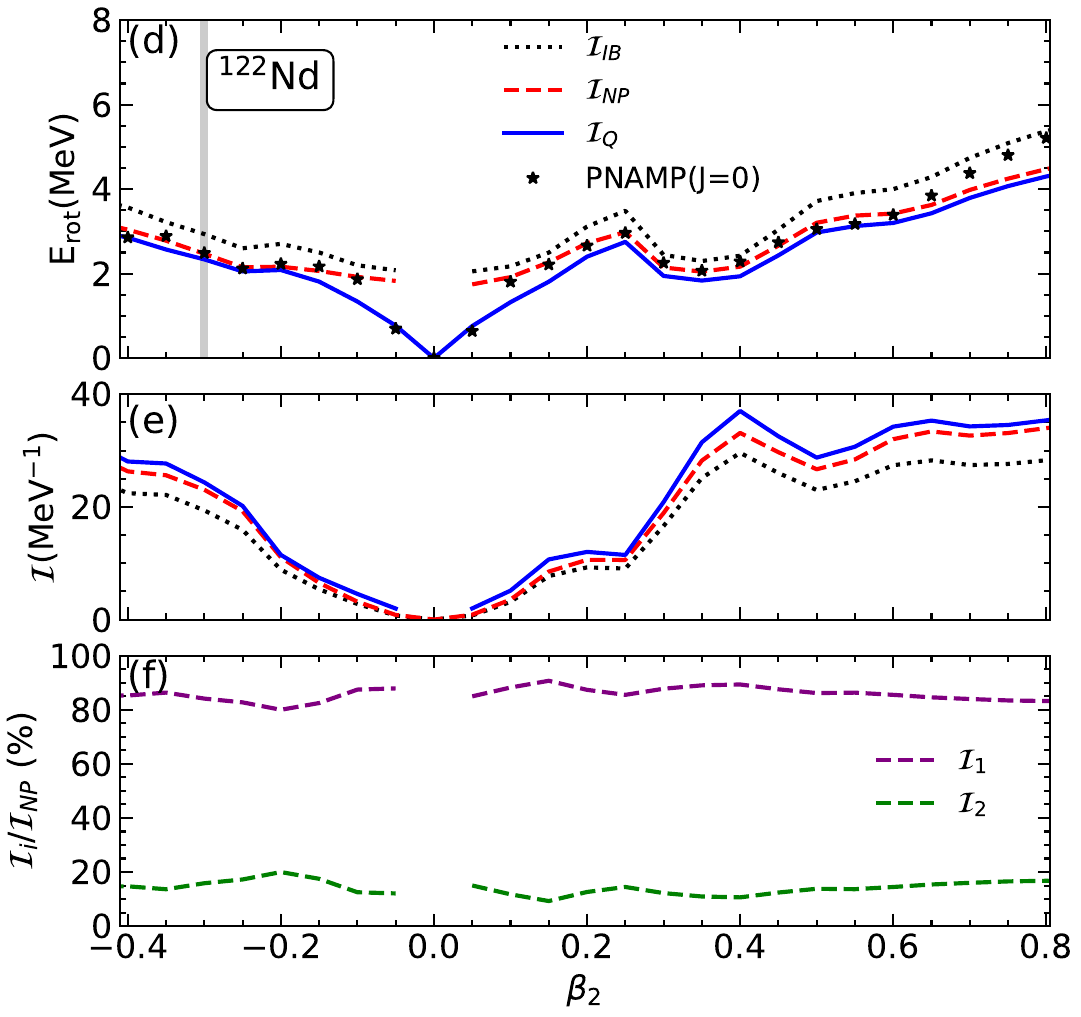}  
	\caption{(Color online) (a, d) Comparison of RCEs and (b, e)  MoIs for \nuclide[26]{Mg} and \nuclide[122]{Nd}, computed using three different methods. The RCEs calculated with $E^{\rm PNAMP}_{\rm rot}$ are included for comparison. (c, f) Ratios $\mathcal{I}_{i=1, 2}/\mathcal{I}_{\rm NP}$ as a function of the deformation parameter $\beta_2$. The mean-field energy minimal state is indicated by a gray vertical line. See the main text for details.}
  \label{fig:26Mg}
\end{figure}
\begin{figure}[tb]
	\centering
	\includegraphics[width=0.9\columnwidth]{ 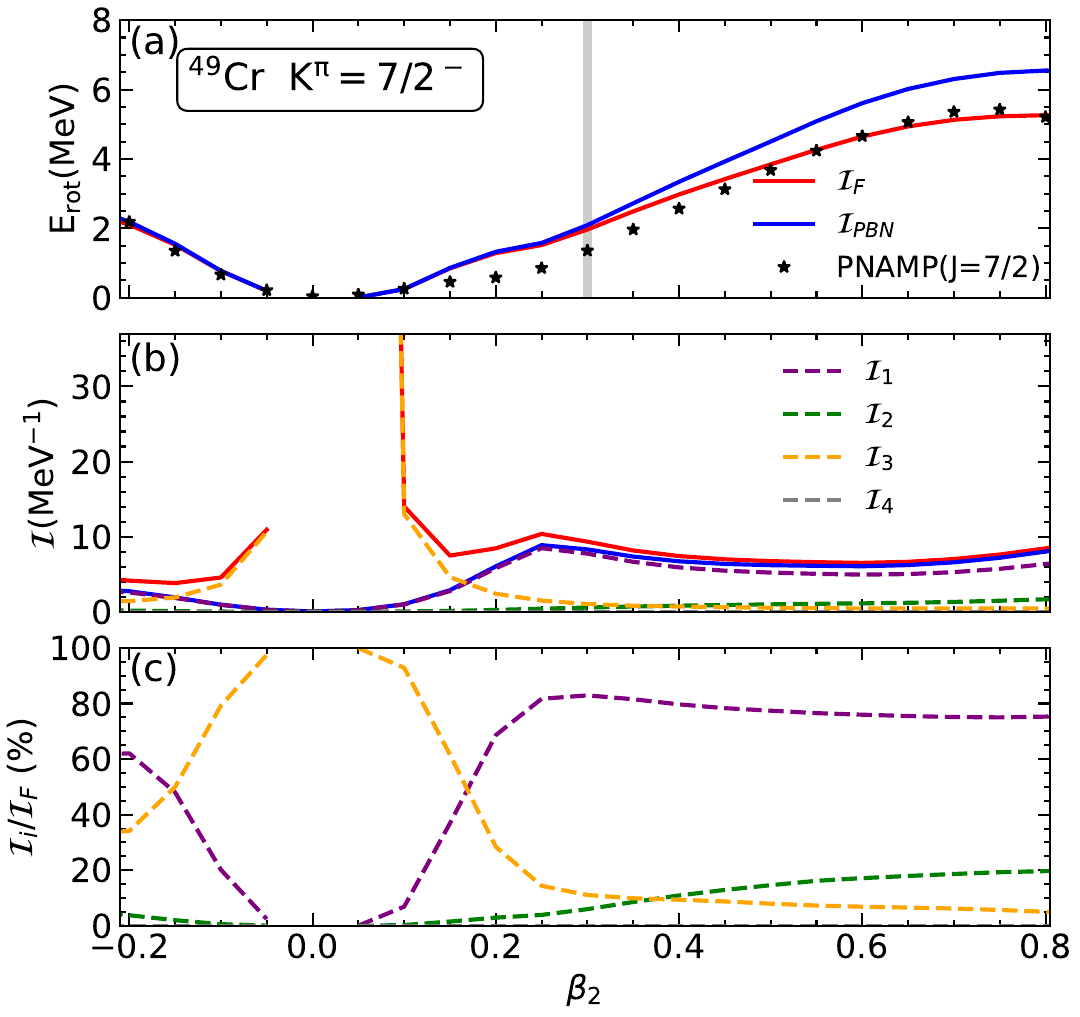}  
	\includegraphics[width=0.9\columnwidth]{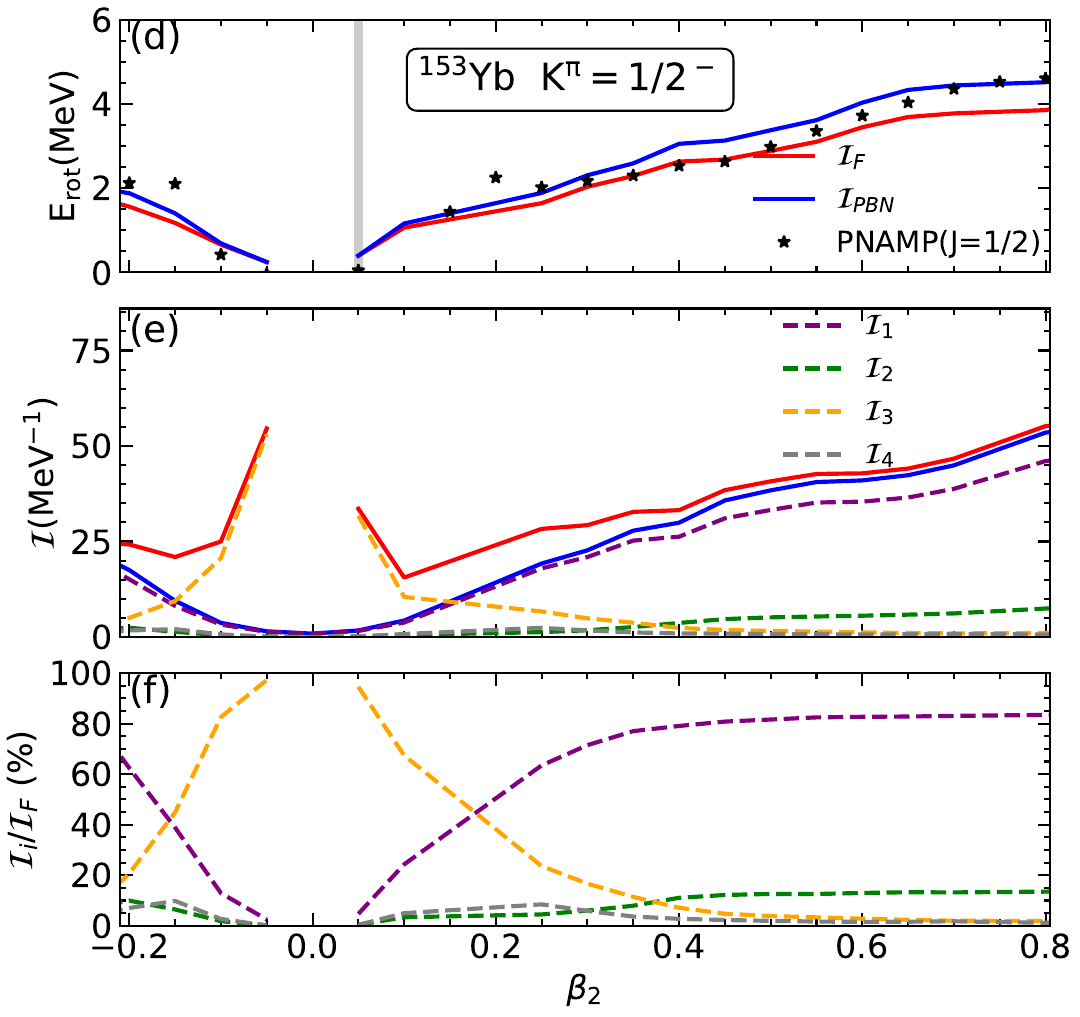}  
 \caption{(Color online) (a, d) Comparison of  RCEs  and (b, e)  MoIs for the lowest quasiparticle state with $K^\pi=7/2^-$ in \nuclide[49]{Cr} and $K^\pi=1/2^-$ in \nuclide[153]{Yb}, computed using two different methods. RCEs calculated with $E^{\rm PNAMP}_{\rm rot}$ with $J=7/2$ and $J=1/2$ for \nuclide[49]{Cr} and \nuclide[153]{Yb}, respectively, are provided for comparison. (c, f) Ratios $\mathcal{I}_{i=1, 2, 3, 4}/\mathcal{I}_{\rm F}$ as a function of the deformation parameter $\beta_2$. The mean-field energy minimal state is indicated by a gray vertical line. } 
  \label{fig:Cr49_72m_ErotMOI}
\end{figure}

\begin{figure}[tb]
	\centering
	\includegraphics[width=\columnwidth]{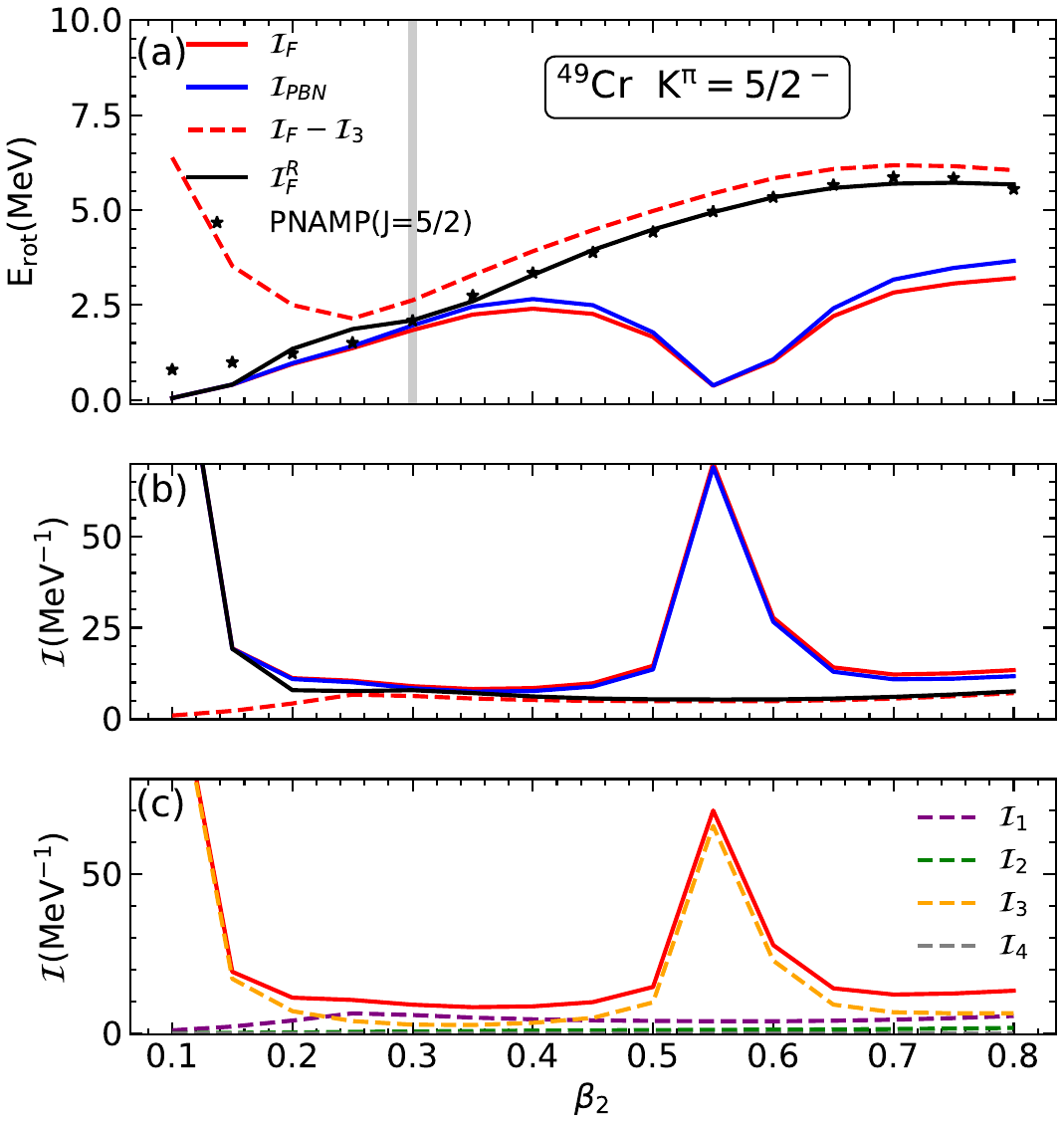}  
	\caption{(Color onlie) Same as Fig.~\ref{fig:Cr49_72m_ErotMOI}, but for the configuration of \nuclide[49]{Cr} with $K^\pi=5/2^-$. The  RCEs obtained using the cranking approximation with four different MoIs, including the regulated MoI $\mathcal{I}^R_F$ (\ref{eq:regulated_MoI}), are compared with the AMP calculation with $J=5/2$.  }
  \label{fig:Cr49_52m_ErotMOI}
\end{figure}

\begin{figure}[tb]
	\centering
	\includegraphics[width=\columnwidth]{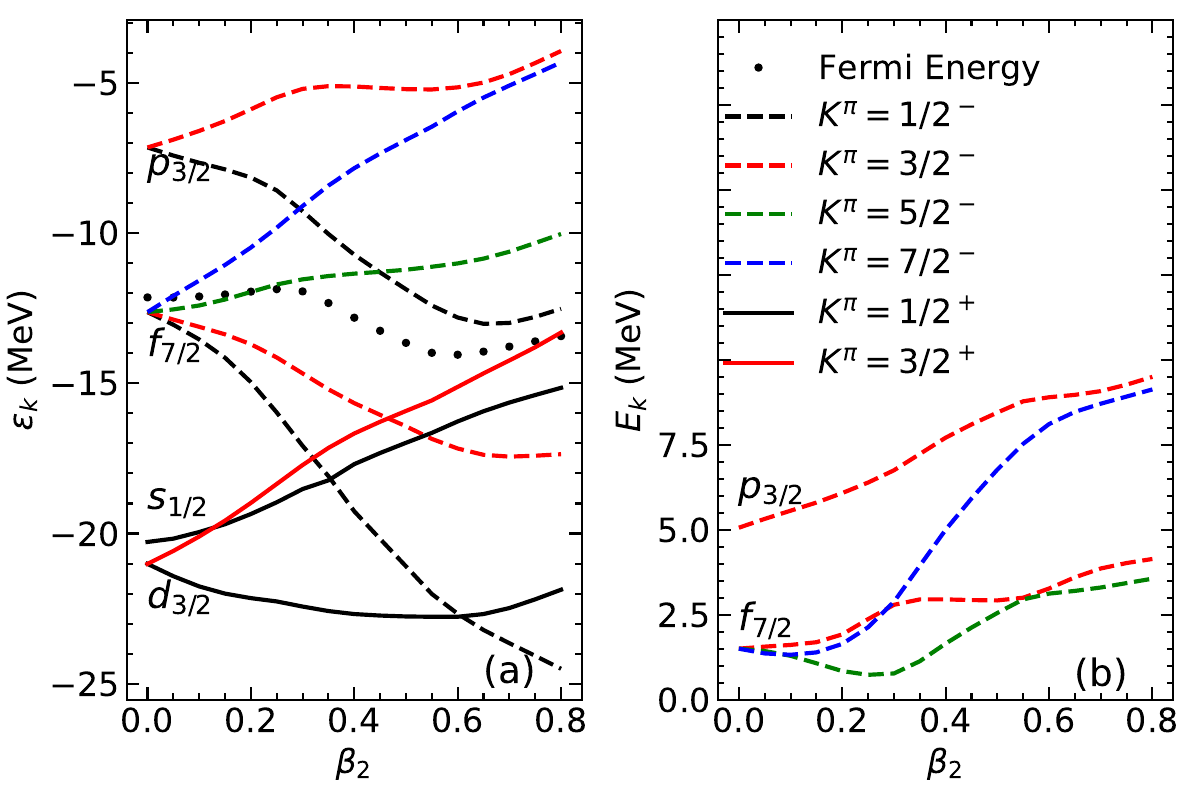}  
	\includegraphics[width=\columnwidth]{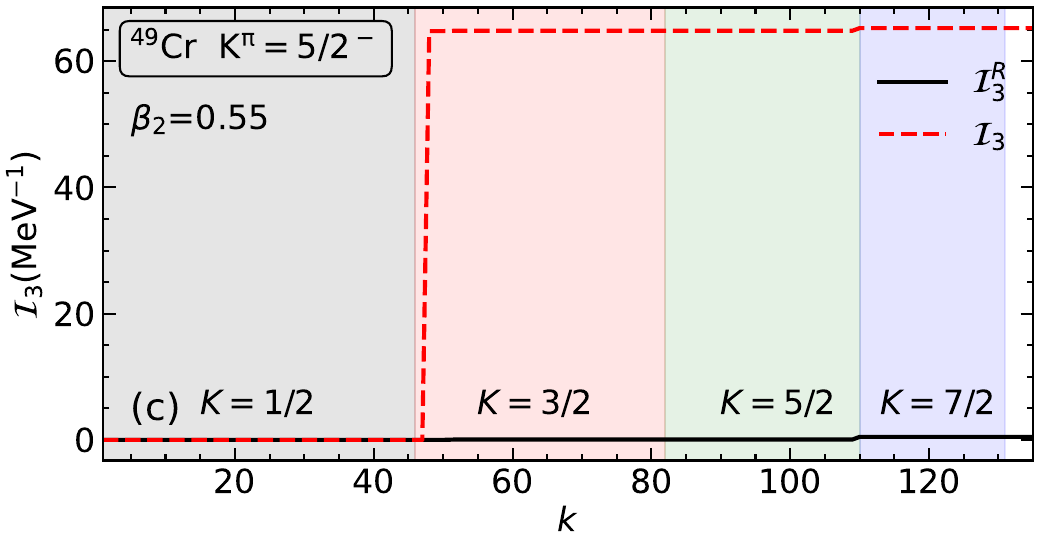}  
	\caption{(Color online) (a) Single-particle energies $\epsilon_k$ and (b) quasiparticle energies $E_k$ in Eq.~(\ref{eq:quasi-particle-energies}) for neutrons in \nuclide[49]{Cr} as functions of quadrupole deformation $\beta_2$. (c) The running sum of $\mathcal{I}_3$ with the index $k$ of single-particle levels, where the blocked orbital is chosen as the one with $K^{\pi}=5/2^-$, originating from the spherical $f_{7/2}$ orbital. }
  \label{fig:Cr49_52m_energy_level_neutron}
\end{figure}

Figure \ref{fig:26Mg} displays the RCEs and MoIs of \nuclide[26]{Mg} and \nuclide[122]{Nd} with the quadrupole deformation parameter $\beta_2$,  
\begin{equation}
  \beta_{2} = \frac{4\pi}{3AR^{2}}\bra{\Phi} r^{2}Y_{20} \ket{\Phi}.
\end{equation}
For comparison, the RCEs by different MoIs are compared. It is shown that the MoI $\mathcal{I}_{\rm IB}$ by the Inglis-Belyaev formula, i.e., the $\mathcal{I}_1$ term, is systematically smaller than the $\mathcal{I}_{\rm NP}$ by the Nilsson-Prior formula (\ref{eq:NP}) due to the contribution of the second term $\mathcal{I}_2$, which increases the MoI overall by about 20\%.  Figure \ref{fig:26Mg}  also shows the RCE by the formula (\ref{eq:pheno_formulas_cranking}), 
\begin{equation}
\label{eq:rot_IB_Q}
  E^{\rm Q}_{\rm rot} =\frac{\braket{\hat J^2}}{2\mathcal{I}_{\rm IB}} \Big[b \tanh(c\abs{\beta_2})\Big] 
  \equiv \frac{\braket{\hat J^2}}{2\mathcal{I}_Q},
\end{equation}
where 
\begin{equation}
     \mathcal{I}_{\rm Q} \equiv  b^{-1}\coth(c\abs{\beta_2}) \mathcal{I}_{\rm IB}.
\end{equation}

According to Ref.\cite{Chamel:2008aa}, the values of $b = 0.80$ and $c = 10$ were determined to reproduce the nuclear mass data, leading to a ratio of $ \mathcal{I}_{\rm Q}/ \mathcal{I}_{\rm IB} \simeq 1.25$. This implies that the MoI $\mathcal{I}_{\rm Q}$ is comparable to $\mathcal{I}_{\rm NP}$ for deformed states, as shown in Fig.\ref{fig:26Mg}. Notably, the RCE for the energy-minimal state using $\mathcal{I}_{\rm NP}$ is closer to the exact AMP calculation than that obtained using $\mathcal{I}_{\rm IB}$. However, it is clear that the RCEs for states near spherical shapes deviate significantly from the exact AMP values, regardless of the formula used for the MoI, highlighting the breakdown of the non-degenerate perturbation theory in the cranking approximation for RCEs. This deficiency can be addressed by introducing the factor $\tanh(c|\beta_2|)$ manually, as shown in Eq.(\ref{eq:rot_IB_Q}). In summary, if the Nilsson-Prior formula is used for the MoI, the parameter $b$ in Eq.(\ref{eq:pheno_formulas_cranking}) is unnecessary, but sill need the factor $\tanh(c\abs{\beta_2})$.

Figure~\ref{fig:Cr49_72m_ErotMOI}  presents the results for the odd-mass nuclei \nuclide[49]{Cr} and \nuclide[153]{Yb}. The block orbitals are selected as the lowest quasi-particle states with $K^{\pi}=7/2^-$ for \nuclide[49]{Cr} and $K^{\pi}=1/2^-$ for \nuclide[153]{Yb}. The full MoI $\mathcal{I}_F$ of an odd-mass nucleus is decomposed into four terms, as shown in Eq.~(\ref{eq:IF}). It is observed that the first term $\mathcal{I}_1$ contributes approximately 80\% for deformed states with $|\beta_2|>0.2$. For weakly deformed states, the MoI is dominated by the third term $\mathcal{I}_3$, which increases rapidly as $\beta_2 \to 0$, while the other three terms approach zero. Consequently, the full MoI becomes large, and the RCE decreases to zero, consistent with the trend observed in the exact AMP results. Notably, the RCEs obtained using the full MoI align excellently with those from the exact AMP. Thus, if the full MoI is used in the cranking approximation formula, there is no need to introduce the free parameter $b$ or the function $\tanh(c\abs{\beta_2})$ for the RCE. However, if the MoI $\mathcal{I}_{\rm PBN}=\mathcal{I}_1+\mathcal{I}_3$ is used, the RCE will generally be overestimated due to the missing $\mathcal{I}_2$ term, which contributes approximately 20\% to the full MoI. The fourth term, $\mathcal{I}_4$, is generally negligible and is nonzero only when the blocked orbital has an angular momentum projection along the $z$-axis of $K=1/2$.

\begin{figure}[tb]
	\centering
	\includegraphics[width=\columnwidth]{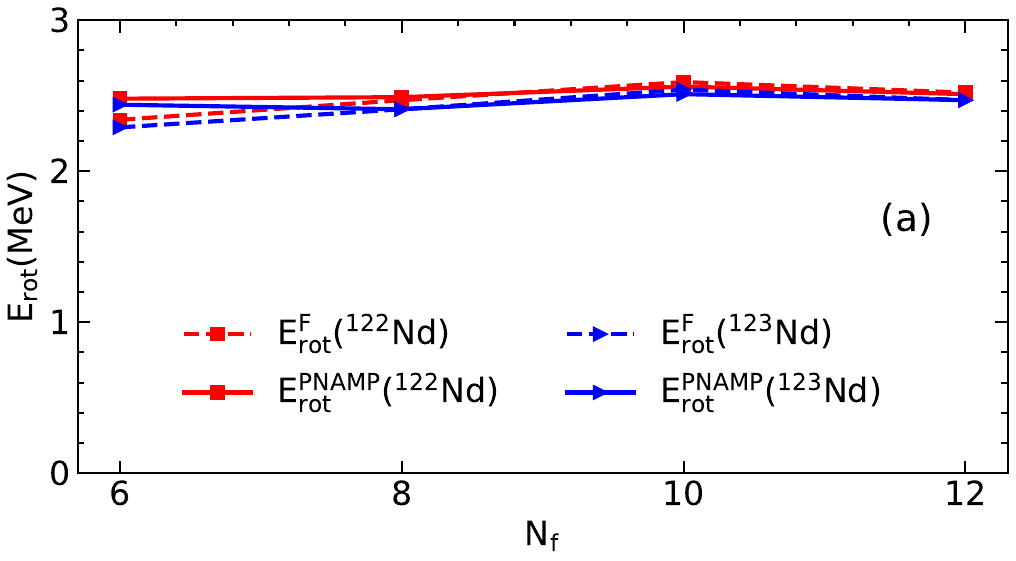} 
        \includegraphics[width=\columnwidth]{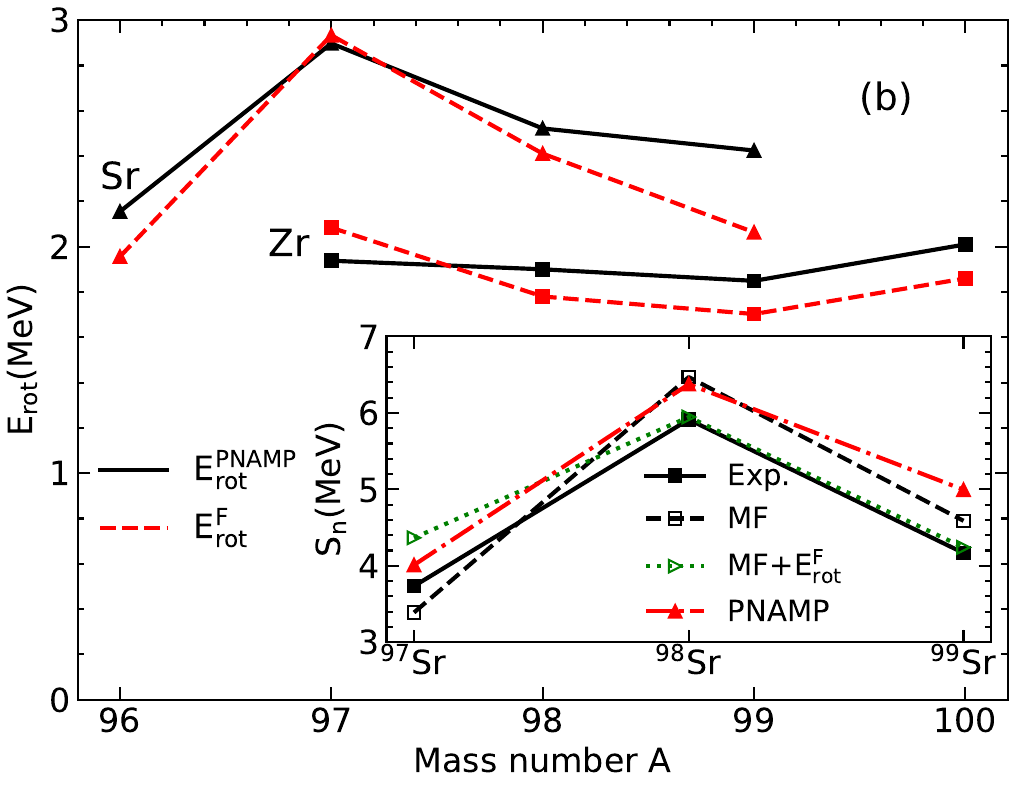} 
	\caption{(Color online) (a) The convergence of RCEs $E_{\rm rot}$ from the exact AMP calculations for particle-number projected states (\ref{eq:PNAMP_rot}) with $\beta_3=-0.3$,  compared to those from the cranking approximation using the Nilsson-Prior MoI  (\ref{eq:NP}) for \nuclide[122]{Nd} and  full MoI for \nuclide[123]{Nd}, as the number $N_{\rm f}$ of HO major shells increases. (b) The RCEs for Sr and Zr isotopes as functions of  mass number $A$, compared between two different calculation methods. 
The inset presents a comparison of the one-neutron separation energies.}
 \label{eq:RCE_Nf_neutron_number}
\end{figure}

\begin{figure*}[tb]
	\centering
	\includegraphics[width=\textwidth]{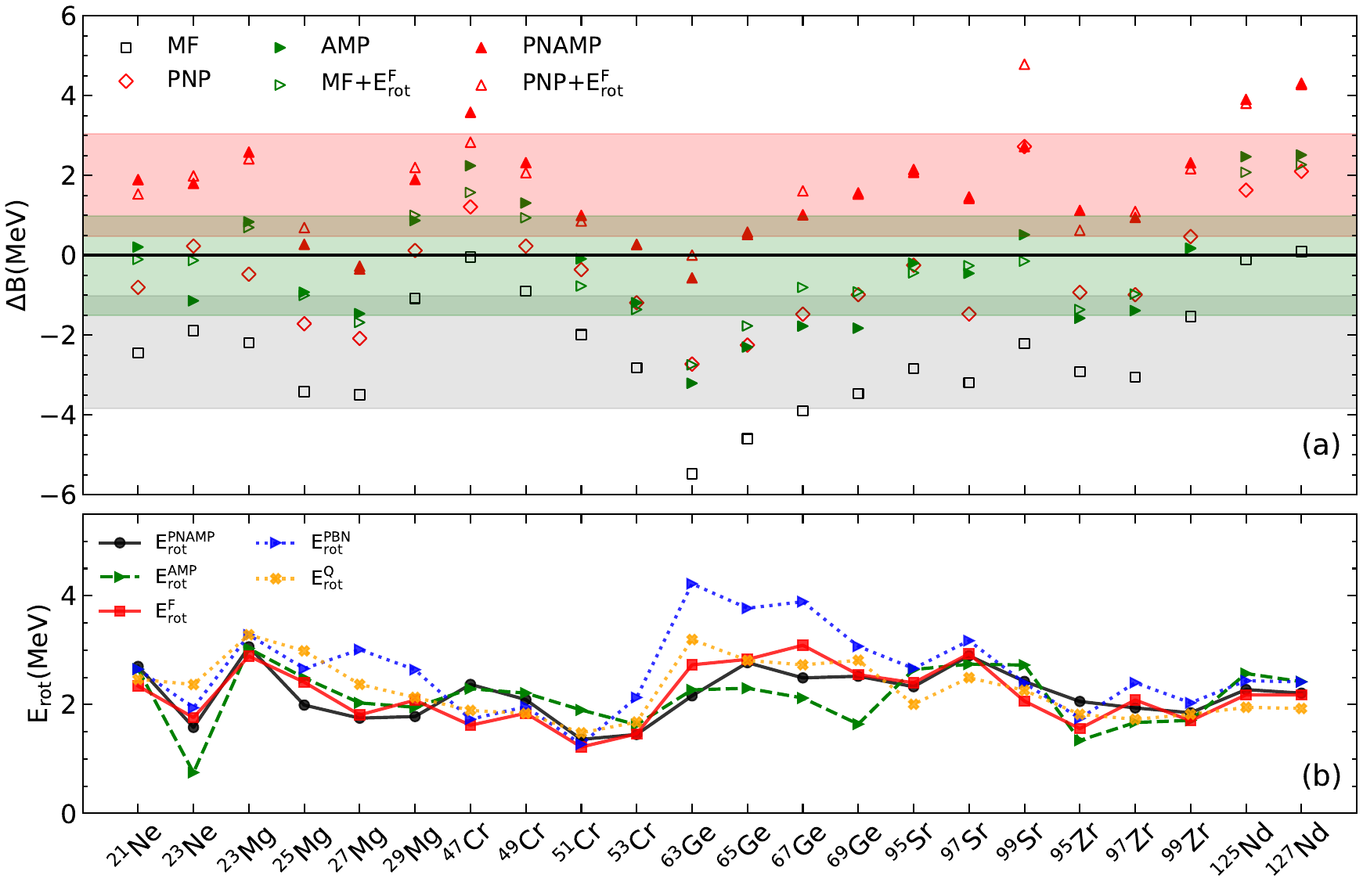}  
	\caption{(Color online) (a) The deviations $\Delta B$ of the predicted nuclear binding energies from corresponding data  for a set of odd-mass nuclei spanning from the light to heavy mass regions, calculated using the PC-F1 EDF. Results are shown for various levels of calculation: pure mean-field (MF), PNP, AMP, MF plus the RCE $E_{\rm rot}^{\rm F}$ from the cranking approximation (MF+$E_{\rm rot}^{\rm F}$), both PNP and AMP (PNPAMP), and  PNP+$E_{\rm rot}^{\rm F}$. The confidence interval for the standard deviation  of $\Delta B$ by the MF, MF+$E_{\rm rot}^{\rm F}$, and PNP+$E_{\rm rot}^{\rm F}$ are indicated with gray, light green and light pink colored areas, respectively.  (b) Comparison of RCEs from different calculations. This includes results from exact AMP for states with PNP ($E_{\rm rot}^{\rm PNAMP}$) and without PNP ($E_{\rm rot}^{\rm AMP}$),  from the cranking approximation using MoI $\mathcal{I}_{\rm PBN}$ and $\mathcal{I}_{\rm F}$, and those by the formula (\ref{eq:rot_IB_Q}), where the quasiparticle vacuum (\ref{eq:odd-mass-qpv}) is used to evaluate $\braket{\hat J^2}$ in Eq.(\ref{eq:expect_J2}).  Nuclei with mass numbers lighter and heavier than $A=90$ are solved using $N_{\rm f}=8$ and $10$, respectively.
}
    \label{fig:nuclear_masses}
\end{figure*}

The MoIs for certain states of an odd-mass nucleus can exhibit unusually large values, as illustrated in Fig.~\ref{fig:Cr49_52m_ErotMOI}. It is observed that the RCEs for \nuclide[49]{Cr} calculated using both the full MoI $\mathcal{I}_{\rm F}$ and $\mathcal{I}_{\rm PBN}$ are similar, but both deviate significantly from the RCE derived from the exact AMP with $J$ projected onto $5/2$. Specifically, the RCE shows a dip around $\beta_2 = 0.55$, which corresponds to the peaks observed in $\mathcal{I}_{\rm F}$ and $\mathcal{I}_{\rm PBN}$. 
Figure~\ref{fig:Cr49_52m_ErotMOI}(c) indicates that the abnormally large MoI around $\beta_2 = 0.55$ is due to the $\mathcal{I}_3$ term. To elucidate this abnormal behavior, Fig.~\ref{fig:Cr49_52m_energy_level_neutron} (a) displays the single-particle energy levels of neutrons in \nuclide[49]{Cr} as a function of the quadrupole deformation parameter $\beta_2$. The level with $K^{\pi} = 5/2^-$, originating from the splitting of the spherical $f_{7/2}$ orbital, is chosen as the blocked orbital. The Fermi energies are close to the energy of the blocked orbital with $K^{\pi}=5/2^-$ when $\beta_2 < 0.3$ and then decrease towards the midpoint between $K^{\pi} = 3/2^-$ and $K^{\pi} = 5/2^-$ as $\beta_2$ increases.  Figure \ref{fig:Cr49_52m_energy_level_neutron} (b) shows the quasiparticle energies of levels with $K^{\pi}=3/2^-$, $5/2^-$, and $7/2^-$. It is observed that the quasiparticle energy $E_k$ of the level with $K^{\pi}=3/2^-$ from $f_{7/2}$ approaches $E_{k_b}$ around $\beta_2 = 0.55$.  Figure \ref{fig:Cr49_52m_energy_level_neutron} (c)  reveals that the running sum of $\mathcal{I}_3$ increases significantly when the contribution of the single-particle level with $K^\pi=3/2^-$ is included, while other levels contribute minimally. This phenomenon is explained by the fact that the level with $K^\pi=3/2^-$ is close to the level with $K^\pi=5/2^-$, causing the denominator $E_k - E_{k_b}$ in Eqs.(\ref{eq:IF}) and (\ref{eq:IPBN})  to approach zero, resulting in an abnormally large value for $\mathcal{I}_3$.

Figure~\ref{fig:Cr49_52m_ErotMOI}(a) shows that excluding the entire $\mathcal{I}_3$ term brings the RCEs of deformed states closer to the values obtained from the exact AMP. However, quantitatively, the RCEs without the $\mathcal{I}_3$ term are systematically overestimated by a factor ranging from 8\% to 25\%, which is consistent with the finding in Ref.~\cite{Scamps:2020fyu}. Additionally, the RCEs for near-spherical states become abnormally large, indicating that simply excluding the $\mathcal{I}_3$ term is not a viable solution.
In constrast, introducing the regulator (\ref{eq:regulator}) results in a regular behavior for the MoI $\mathcal{I}_F^R$ for the $K^\pi=5/2^-$ configuration, which is similar to that observed for $K^\pi=7/2^-$ in Fig.~\ref{fig:Cr49_72m_ErotMOI}. Specifically, the RCEs using the regulated MoI are much closer to the values obtained from the exact AMP, as shown in Fig.~\ref{fig:Cr49_52m_ErotMOI}(a).

Figure~\ref{eq:RCE_Nf_neutron_number}(a) compares the RCEs of the even-even nucleus \nuclide[122]{Nd} and the odd-mass nucleus \nuclide[123]{Nd} from the cranking approximation and exact AMP calculations. It is observed that the RCEs obtained from the cranking approximation are in excellent agreement with the discrepancy less than 0.2 MeV. Moreover, the RCE values remain within $2-3$ MeV and show minimal variation with the number $N_{\rm f}$ of HO major shells, indicating that the RCEs are relatively insensitive to $N_{\rm f}$. Figure~\ref{eq:RCE_Nf_neutron_number}(b) shows the RCEs for Sr and Zr isotopes, which change smoothly between $2-3$ MeV as a function of the mass number $A$.  Among these isotopes, the largest discrepancy about 0.5 MeV between the RCEs obtained from the two methods is observed in \nuclide[99]{Sr}. In particular, the RCE of \nuclide[97]{Sr} is evidently larger than neighboring even-even nuclei, and thus have a significant impact on the predicted one-neutron separation energies.

Figure~\ref{fig:nuclear_masses} compares the deviations of predicted binding energies from corresponding data and RCEs from various calculations for a set of deformed odd-mass nuclei ranging from \nuclide[21]{Ne} to \nuclide[127]{Nd}, where the quadrupole deformation parameters $\beta_2$ of their energy-minimal states range from 0.20 to 0.52. Figure~\ref{fig:nuclear_masses}(b) shows that the RCEs across all methods fluctuate around 2.0 MeV within a range of $1-4$ MeV, depending on their $\beta_2$ values. The RCEs obtained by different methods may differ by more than 1.0 MeV. Notably, the RCEs from the cranking approximation using the full MoI $\mathcal{I}_{\rm F}$ are in close agreement with those from exact AMP calculations. In contrast, using the MoI $\mathcal{I}_{\rm PBN}$ systematically overestimates the RCEs. Furthermore, the RCEs of $E^{\rm PNAMP}_{\rm rot}$ differ slightly from those of $E^{\rm AMP}_{\rm rot}$, with discrepancies generally less than 0.5 MeV. In particular, one can see that the discrepancy between $E^{\rm AMP}_{\rm rot}$ and $E^{\rm F}_{\rm rot}$ is comparable to that between $E^{\rm AMP}_{\rm rot}$ and $E^{\rm Q}_{\rm rot}$. It indicates that the phenomenological formula (\ref{eq:rot_IB_Q}) works rather well for odd-mass nuclei as well.

Figure~\ref{fig:nuclear_masses}(a) demonstrates that the   binding energies of pure mean-field calculation are systematically lower than the data. Including the RCEs $E^{\rm F}_{\rm rot}$ reduces the rms error compared to experimental data from 2.77 MeV to 1.29 MeV, which is close to the 1.57 MeV achieved with $E^{\rm AMP}_{\rm rot}$. This is consistent with the finding in even-even nuclei~\cite{Zhang:2013FP}. For comparison, we also evaluate the deviations of predicted binding energies by including the RCEs  $E_{\rm rot}^{\rm Q}$ and find the rms error 1.20 MeV. It demonstrates again the validity of the phenomenological formula (\ref{eq:rot_IB_Q}) for odd-mass nuclei. In contrast, the rms error from the PNAMP calculation becomes 2.15 MeV. This indicates that the inclusion of the DCEs from both PNP and AMP systematically overestimates nuclear binding energies, suggesting that some of the DCEs may have already been incorporated during the optimization of the EDF parameter set at the mean-field level.

\section{Summary}
\label{sec:summary}

In this work, we conducted a benchmark study of rotational correction energies (RCEs) for a set of quadrupole-deformed odd-mass nuclei in the cranking approximation, comparing them to exact angular-momentum projection (AMP) results within a covariant DFT framework. We evaluated cases with and without particle-number projection. Our findings reveal that, for certain configurations in odd-mass nuclei, the RCEs can become abnormally small due to divergence in the moment of inertia (MoI). To address this issue, we introduced a regulator for the divergent terms in the MoI and validated the use of the regulated MoI through exact AMP calculations. Our results show that RCEs computed using the full MoIs in the cranking approximation closely match those from exact AMP for configurations with good nucleon numbers. The inclusion of RCEs significantly improves the description of the binding energies of these odd-mass nuclei, reducing the rms error from 2.77 MeV to 1.29 MeV. However, nuclear binding energies are systematically overestimated by about two MeV when additional dynamical correlation energies from particle-number projection are included. This suggests a need to readjust the parameter set of the underlying EDF at the beyond mean-field level.

\section*{Acknowledgments} 

We extend our thanks to C.F. Jiao, Y.X. Zhang, and C. Zhou for their insightful discussions, and to S.Q. Zhang for carefully reviewing the manuscript and suggestions. This work is partially supported by the National Natural Science Foundation of China (Grant No. 12375119), the Guangdong Basic and Applied Basic Research Foundation (2023A1515010936), and the Fundamental Research Funds for the Central Universities at Sun Yat-sen University.

\newpage 

   
%

\end{document}